# Stochastic Thermodynamics and Dynamics:
# A Tail of Unexpected

**Maria K. Koleva**

*Institute of Catalysis, Bulgarian Academy of Sciences,*
*1113 Sofia, BULGARIA*
E-MAIL: *mkoleva@bas.bg*

The problem of the insensitivity of the macroscopic behavior of any thermodynamical system to partitioning generates a bias between the reproducibility of its macroscopic behavior viewed as the simplest form of causality and its long-term stability. The overcoming of this controversy goes through certain modification of the dynamics that involves self-assembling of the boundary conditions. Subsequently the proposed approach justifies parity between the increase and the decrease of the entropy which provides the ground for holistic understanding of the thermodynamical systems through launching their ability to transmit and create information that is sensitive to coherent functioning of self-assembled logical landscapes. The obtained sensitivity gives the advantage of this new approach compared to that of Shannon. According to his definition, the information depends only on the overall probability for realization of a given state(s) and thus it does not distinguish between functionally different states provided the overall probability for the realization of each of them is equal.

## Introduction

The growing interest in the stochastic thermodynamics has been provoked by its recent application to informational, social and ecological systems. Despite its considerable potential for explanation of various phenomena the diversity of dynamical rules strongly challenges its fundamental postulate inferring separation of the time scales such that to render the insensitivity of the macroscopic behavior to the particularities of the dynamics of the constituents. The assumed insensitivity makes possible the association of time-independent frequencies with the development of every macroscopic fluctuation. Therefore the time average on every macroscopic time scale becomes uniquely determined and time-translational invariant. The importance of the latter property is that it provides the reproducibility of the events, both deterministic and stochastic. If considered as broken time-translational invariance, the manifestation of each event would ambiguously depend on the moment of its realization even when the conditions for its performance remain unchanged. This type of reasoning is in conflict with the reproducibility viewed as the simplest form of causality and by opposing straightforward cause-effect relationships it conditions the crucial importance of establishing the circumstances which validate the aforementioned postulate. As a primary concern for finding solution of this disparity, we will demonstrate that the general condition for its validation confronts both the classical and the quantum dynamics.

In order to outline this problem we will start examining the general prerequisite of the target invariance: the extensive macroscopic variables (e.g. concentration, temperature) are to be invariant with respect to the partitioning of the system; at the same time the intensive macroscopic variables must be additive regardless to the way the partitioning is made. It is beyond doubt that the independence from the partitioning justifies the uniqueness of the averaged product and its independence from the details of the dynamics of the constituents. If otherwise, i.e. in the case of any dependence on the partitioning, the macroscopic variables depend on the choice of the auxiliary reference frame selected to define the concrete way of partitioning. So far it has been taken for granted that the avoidance of this problem is justified by the assumption that since the macroscopic variables vary much slowly than the fast dynamical ones, the averaging over the entire phase space (ensembles in the classical case) would take place at every location in the system considered. Therefore the same averaged probability is assigned to every constituent. The elusive point, however, resides in the lack of any comprehensive argument whether the averaging protocol uniformly converges to an invariant for the spatio-temporal location value. The importance of this issue lies in the fact that it acts as a criterion that discriminates two fundamentally different conducts: whilst any non-uniform convergence makes the macroscopic behavior critically dependent on the choice of the auxiliary reference frame and thus provokes conflict with time-translational invariance, the uniform convergence implements elimination of any such dependence. This subject is discussed in the next section where key arguments for the non-uniformity of that convergence are proposed. Moreover, we will prove that the core of the problem is deeply rooted into the dynamics. We will go further and will demonstrate that the modification of the dynamics, recently proposed by us [1,2], successfully overcomes the above difficulties at the expense of approaching the aforementioned postulate by means of replacing the averaging over the phase space with spontaneous synchronization available through the considered modification of the dynamics. Moreover, it turns out that the developed approach automatically justifies the launch of the thermal equilibrium as universal dynamical process such that in every single collision the dissipation goes from the species of larger energy to the species of lower energy. Still, the major merit of the proposed modification is its consistency with the recent reformulation of the thermodynamics [3] that opens the ground for holistic understanding of the thermodynamical systems through launching their ability to transmit and create information through sensitivity to coherent functioning of self-assembled logical landscapes.

## 1. The Conflict

As discussed in the Introduction, the generation of a conflict between the thermodynamics and the time-translational invariance crucially depends on the answer to the question whether the uniform convergence is a generic property of the partitioning. The aim of the present section is to prove that there is no uniform convergence and to demonstrate that the problem is to be traced down to the dynamics. In order to make this proof more evident, we will apply our argumentations to the process of relaxation of adsorbed/absorbed species. The generalization of the problem for the collisions in gases is considered in sec.3.

The adsorption/absorption is process such that species confined by the surface/volume relax to the ground state of potential wells situated at certain locations on a surface/volume. The transitions through which the relaxation proceeds, dissipate trough excitation of those collective surface/volume modes which match certain specific for the transition and the mode conditions called hereafter resonance conditions. The excited collective modes, in return, act as a perturbation on the potential well and thus modify the resonant conditions. Since these perturbations are maintained by the current spatio-temporal configuration, the question is whether they drive any dependence of the probability for relaxation on its morphology. To find out, let us have a look on the energy of the excited modes in an arbitrary location and at arbitrary instant:

$$J_{exc}(x,t) \approx \sum_i \int_\omega \int_\tau d\omega d\tau A(\omega) \exp\left(i\omega(t - \tau_i(x,t)) - i\vec{k}(\omega) \bullet (\vec{x} - \vec{\mu}_i)\right) \quad (1)$$

where $\omega$ is the frequency of an excited mode; $A(\omega)$ - its amplitude; $\vec{k}(\omega)$ is the dispersion relation of the mode; $\vec{\mu}_i$ is the location of the $i-$th potential well; $\tau_i(x,t)$ is the delay of the mode that travels from the $i-$th potential well.

The procedure that examines the convergence of the partitioning protocol is well established and involves the following steps: (i) considering an window of arbitrary size $(\varsigma, \theta)$ located at the point $(x,t)$, the specification of the margins of integration over the frequencies of the excited modes $\omega$ and the delays $\tau$ that fit the window considered is to be set; (ii) the second step is establishing whether the averaged over that window dissipated energy converges to an invariant for the system value. The intrigue in this procedure is the revelation of a so far overlooked relation between the frequencies and the delays. Indeed, as explained above, any change of the dissipated energy is carried out by those transitions that match the specific for the dynamical process resonance conditions which provide correspondence with the excited collective modes. On the other hand, the excited modes perturb the potential wells and thus modify the resonance parameters. Bearing in mind the general rule of the dynamics that the resonance conditions are highly specific to the process considered, the independence of the delays from one another makes the dissipated energy fine-tuned to the spatio-temporal location which in turn provokes high sensitivity of the averaged product to the window parameters. That is why the averaging drives amplification of even infinitesimal differences in the dissipated energy and so intensifies its non-uniformity. Moreover, the joint action of the fine-tuning and the produced by it non-uniformity brings about additional "scattering" of the delays which results in further escalation of the non-uniformity. Hence, as follows from (1), the explicit dependence of the averaged product on the window parameters $(\varsigma, \theta)$ and its location $(x,t)$ illuminates the assertion that the non-uniformity is straightforwardly related to the choice of the auxiliary reference frame that serves to describe the way of partitioning.

Yet, it seems that the question about the uniformity of the convergence of the partitioning is a matter of trivial mathematical argumentation because the postulated in the stochastic thermodynamics large gap between the macroscopic and dynamical time scales suggests that there is "enough time" for performing all dynamical configurations and so the averaged probability converges to the same value at every location. That is why at every location, the integration over all available delays is executed along with the integration over all available frequencies. Though this protocol eventually renders the averaged probability invariant, the current non-uniformity drives a conflict with the stability of the system. Indeed, the non-uniformity permanently propels sharp discontinuities in the spatio-temporal configurations that on exceeding the local thresholds of stability bring about irreversible local defects whose unrestrained development would ultimately result in the system breakdown.

Another argument in favor of the high non-triviality of the subject considered is the relation between the issue of the time scale separation and the velocity anzatz which states that the velocity of transmitting matter/energy/information through every media is always bounded. The account for the boundedness of the velocity of every motion makes the time for performing all dynamical configurations (time of convergence) subject to the size of the system. If so, the response to every perturbation would be intensive variable which renders the stability of the system subject to its size – an obvious conflict with our experience.

To summarize, the issue about the uniform convergence of the averaging protocol brings together the time-translational invariance and the stability in a startling puzzle – along with the stability of the system considered, the unrestrained development of local instabilities violates the time-translational invariance of its macroscopic behavior. This gives rise to the question whether the time-translational invariance and the stability are somewhat coupled. In the next section we will demonstrate that they indeed are counterparts because the general approach [2] that provides long-term stability automatically justifies time-translational invariance of the macroscopic behavior. The matter is highly non-trivial since it concerns the fundament of the dynamics – as the considerations in the present section convey, the core of the problem is the fine-tuning of the dissipation to the current morphology, a fine-tuning that originates from the necessity of meeting highly specific for every process resonance conditions.

## 2. The Edge of Confinement

The first task of this section is to delineate the setting where the elimination of the non-uniformity links the stability and the time-translational invariance into a self-consistent framework. For this purpose, let us first note that the physical carrier of the non-uniformity are the extra-gradients produced by the unrestrained development of the local instabilities. The stability of any system, considered as a single object, imperatively requires synchronous smoothing of all extra-gradients. Note that the fulfillment of this condition makes the stability independent from the size of the system. The intrigue of this situation is that together with the stability, the same condition (synchrony of smoothing) provides lessening of the non-uniformity and so contributes to the independence of the averaged product from the partitioning. However, since the synchronous smoothing implies that all the instabilities fade away at the same time, it requires certain relation between the strength of each local instability and the rate of its weakening. Yet, since the strength of the instability and the rate of its weakening are explicitly related to binding and dissipation correspondingly, the problem is unavoidably reduced to the underlying dynamics. It is obvious that the target synchronous smoothing is provided if and only if the resonance conditions are universal, i.e. in the sense that they are insensitive to the particularities of the dynamical process. However, this viewpoint runs the following conundrum: on the one hand, the fundamental merit of the specificity of the resonance conditions is that it guarantees preserving the identity of the species during interaction. On the other hand, as we demonstrated above, the avoidance of the unrestrained development of local instabilities requires universality of the resonance conditions which, however, automatically discards the identity of the species. Therefore our goal is to demonstrate that the way to reconciliation of these highly contradictive subjects is a new approach to the dynamics. For this purpose, let us first consider the role of the dynamics in providing the stability.

A general contemplation of the dynamics is to consider the characteristics of the confinement of a ***single*** species in a certain area of *apriori* settled ***steady*** boundary conditions. Viewed from the prospective of the stability, the confinement of a limited number species in an area/volume of finite size is the major implement for preventing accumulation of arbitrary amount of energy/matter in any location. Thus it defines thresholds for the matter/energy to be incorporated (adsorbed/absorbed) in a media so that not to destabilize it. Further, since the confinement reduces any long-range interaction to a short-range one, it makes the species in different locations independent from one another objects. In turn, this justifies the well-known phenomenon of saturation to be viewed as threshold of the local concentration whose surpass yields destabilization. On the other hand, however, as shown above, the surpass seems unavoidable because the specific resonance conditions do not allow smoothing of the dissipated energy over the entire system. Thus, the core of the problem is the paradigm about the permanency of the boundary conditions because it is it that renders the resonance conditions non-generic. In order to solve the problem we replace the idea of apriori setting of the boundary conditions with the idea of self-assembling of the adsorption potential wells and their boundary conditions. This task needs new viewpoint on the dynamics since the all known so far approaches operate at apriori set steady boundary conditions. However, the fundamental flaw of these approaches is that they consider action of more than one perturbation always as a linear superposition whose drawback, however, is that it allows accumulation of arbitrary large amount of energy/matter in every location and thus confronts the stability. In order to avoid accumulation of arbitrary amount energy /matter we replace the idea of linear superposition with the idea of boundedness. The latter implies that the energy/matter in every moment and in every location stays permanently bounded. Our next task is to prove that this idea gives rise to universal relation between binding and dissipation which automatically eliminates the non-uniformity. We will demonstrate that the elimination of the non-uniformity happens at zero binding energy and after that the further relaxation proceeds according to the conventional dynamics. Our next task is to delineate the realm of boundedness and the realm of linear superposition. We will demonstrate that their complementarity provides both the elimination of non-uniformity and preserving the identity of the species and the interactions among them.

The crucial demarcation line between the realm of linear superposition and the realm of boundeness is the edge of confinement, i.e. the zero binding energy. Indeed, at zero binding energy each perturbation, regardless to its strength, produces radically different effect: it causes either confinement or free movement. The crucial point however, is that this effect is universal because the details of the perturbation are unimportant: at zero binding energy de Broigle wavelength is arbitrarily large which makes the interacting objects not to "feel" the particularities of their identity. Thus, the interactions at the edge of confinement are universal in the sense that they are insensitive to the identity of the interacting species and the nature of the interactions. Their universality is conditioned only by the requirement about the boundedness of the accumulated energy so that to keep the system within its thresholds of stability. The question now is whether the characteristics of the relation between the binding and the dissipation subjected to the boundedness are also universal and if so, how they contribute to the solution of the problem about the non-uniformity?

The study of the problem starts with noticing that at the edge of confinement the influence of the excited collective modes to the states of zero binding energy cannot be treated as perturbation. We suppose that the excited modes essentially contribute to the modification of the boundary conditions and thus induce new transitions which dissipate again through excitation of collective modes. In turn, the collective modes spread the dissipated energy throughout the system. Obviously, the process stops when the dissipated energy is evenly distributed throughout the entire system and all relaxing species are in the same state. A circumstance that makes this feedback universal is that the smoothing out of the identity of the species and the media renders the collective modes to be "seen" as acoustic-phonon type excitations, i.e. as excitations whose dispersion relation does not depend on the particularities of the media but on a single parameter (sound velocity). Then, as it is easy to be seen from (1) the linear dispersion relation eliminates the explicit dependence on the auxiliary coordinates. Indeed, taking into account the linear dispersion relation:

$$k(\omega) = \frac{\omega}{c} \qquad (2)$$

$J_{exc}(x,t)$ becomes:

$$J_{exc}(x,t) \approx \sum_i \int\int_{\omega\,\tau} d\omega d\tau \exp(i\omega\tau g(x,t)) P(\omega) G(\tau) \qquad (3)$$

where $g(x,t)$ accounts for the current configuration of the non-uniformity; $P(\omega)$ is the distribution of the transition energies ( dissipated energy) and $G(\tau)$ is the distribution of the delays. The intrigue now is to demonstrate that $P(\omega)$ and $G(\tau)$ are not independent from one another but are joined in a single universal relation that ultimately yields evening of the dissipated energy. To prove this assertion let us note that the general prerequisite of the lost identity at the edge of confinement is the lack of any steady symmetry of the boundary conditions. In our previous work [2,&5.5] we have proved that the statistical properties of the binding at the edge of confinement are identical and insensitive to the particularities of the boundary conditions. Moreover, we have proved that the eigenvalues are complex and are uniformly distributed at the unit circle. This immediately yields strong parity between the binding energy (real part of an eigenvalue) and the rate of dissipation (the imaginary part of that eigenvalue) – the larger the binding energy is the slower its dissipation is. Further, considering that the larger instabilities are more weakly bind, it becomes obvious that the larger instabilities dissipate faster than the smaller ones. Therefore, the tendency is the evening of the rates and amplitudes of all instabilities so that eventually their elimination to result in uniform distribution of all dissipated energy throughout the entire system and all the species to have the same characteristics. We call this process spontaneous synchronization in order to distinguish its operational protocol from averaging over the dynamical variables. The greatest merit of the spontaneous synchronization is that the independence from the partitioning is achieved at permanent respecting the stability of the system. The further relaxation proceeds according to the conventional dynamics.

### 3. Atom as a Network. Thermal Equilibrium

So far we have considered the effect of the modification of the dynamics at the edge of confinement only for the system where species interact with a media. However, the universality of our approach requires justification of its successful operation in systems where the only interactions are collisions among the constituents. It is well known, that the dissipative mechanisms proposed so far are model-dependent and does not yield stable evening of the dissipated energy throughout the system and thus fail in providing the insensitivity to partitioning as a generic property. That is why the application of our approach to the case of collisions is a crucial test for its universality.

The major difference with the previous case is that now the de Broigle wavelength if finite because the species move ballistically between collisions. Then its size defines the area within which the interactions proceed under the condition of lost identity. Indeed, inside the latter area, the structure of the species is not well discerned and thus each species "feels" any other as a structureless object. Further, taking into account the postulate about the boundedness of the rate of exchanging matter/energy/information, this structureless object must be considered rather as a universal network, i.e. its only active elements are the acoustic-phonon type collective excitations (lost identity). Then, the exchanged energy in that process is:

$$J_{exc}(x,t) \approx \int d\tau \int_0^{Vk} d\omega \exp\left(i\omega(t-\tau_i) - i\frac{\omega}{V} \bullet (\vec{x}-\vec{x}_i)\right) \qquad (4)$$

where $V$ is the velocity of the species. Thus, the process of dissipation of the kinetic energy to the internal degrees of freedom (network) is reduced to the same type spontaneous synchronization as considered in the previous section. The peculiarity, however, is that the utilization of the dispersion relation of a de Broigle wave:

$$k = \frac{\omega}{V} \qquad (5)$$

makes evident that the larger the velocity of the species is the more energy dissipates from the kinetic energy to the internal degrees of freedom, i.e. to the network. Besides, since the mechanism of this dissipation is universal and independent from the identity of the species, we may well proclaim that it is the implement of launching thermal equilibrium, i.e. equilibrization of the kinetic energy of all the species. Note, that since the proposed by us mechanism is bound to provide the stability of the system, the thermal equilibrium is a manifestation of a long-term stable behavior of a many-body system as a single object. This justifies the key difference between our approach and the conventional thermodynamics, namely: the latter considers the thermal equilibrium as a prerequisite of establishing thermodynamical equilibrium viewed as the state of maximum entropy. Besides, since the edge of confinement is additively decomposed to conventional dynamics, and under the supposition of self-assembling of the boundary conditions, the final macro-state can be any steady state that is long-term stable but **not** necessarily the state of maximum entropy! This result opens the door for reversibility of the processes accompanied by entropy change, i.e. it substantiates parity between the increase and the decrease of the entropy and thus justifies them as physical ground for creating and transmitting information. Note, that the reproducibility of the information requires return in the initial state, a circumstance that imperatively demands parity between both increase and decrease of the entropy. This result is in total accordance with our recently proposed major modification of the thermodynamics [3] whose priority is to set the grounds for many-body systems not only to transform work into heat but to transmit and create information as well. A very important point of the new thermodynamics is the reformulation of its second law because the generation of any information happens at the expense of entropy decrease which apparently violates the assertion that on the approach to the equilibrium the entropy always monotonically increases.

The self-consistency of the proposed modification of the thermodynamics [3] and the present considerations is conveyed by the same general principle, that of the boundedness. Indeed, the proposed modification of the thermodynamics is grounded on the boundedness as a general principle that provides long-term stability. In addition to that we prove now that the same principle provides not only the stability of a system but the reproducibility of its macroscopic behavior as well.

## 4. Entropy, Information and Functional Diversity

The aim of the present study is to demonstrate that the problem concerning the insensitivity to partitioning goes far beyond the trivial mathematical argumentation and leads to new unexpected results. The non-triviality of the matter is that it involves into a complex interplay the fundamental issues such as reproducibility of the macroscopic behavior and ability for creating information with the question of the stability of a system viewed as conditions for self-organization of the constituents into a single long-term stable object. Furthermore, we proved that successful resolving of the problems involves certain modification of the dynamics which in turn makes the formation of the thermal equilibrium a universal dynamical process insensitive to the particularities of the dynamics of the constituents. Consequently, it establishes parity between the increase and the decrease of the entropy which launches the ability of the thermodynamical systems not only to transform work into heat but to create and transmit information as well. Moreover, it opens the door for holistic approach to the matter by means of defining information so that to be sensitive to the functional morphology of a self-assembled logical landscape. We have achieved this goal by defining the information through the properties of the power spectrum of the state that represents a logical unit [3]. The functional morphology of the logical landscape, described by eq.(4) in [3], is mapped onto the properties of the discrete band of the power spectrum so that any change in it yields change in the structure of the discrete band. The obtained sensitivity provides the major advantage of our definition of information compared to that of Shannon. According to his definition [4], the information depends only on the overall probability for realization of a given state(s) and thus it does not distinguish between any functionally different states provided the overall probability for the realization of each of them is equal. Yet, the fundamental difference between our approach and that of Shannon is in defining the physical origin of the information and the noise: on the contrary to Shannon who allows both the noise and the information to be created by the same stochastic process, we provide comprehensible physical ground for the discrimination between information and noise through associating the information with causal relations and deterministic processes and characterizing the noise by the stochastic processes. The lack of any physical discrimination between the information and the noise gives rise to complementarity between the Shannon information and the entropy viewed as a measure of disorder. In turn, the obtained complementarity makes the information a quantitative characteristic whose value is subject to uncertainty determined by the entropy. The decisive drawback of this duality, however, is that it generates a conflict with the time-translational invariance as long as the non-zero entropy renders non-zero probability for deviation from any long-term steady behavior regardless to whether the conditions for reproducibility are the same or not. The conflict is best elucidated by the paradox that it generates, namely: the considered complementarity renders the equilibrium, viewed as the state of maximum entropy, to be the state whose behavior would be predicted with the lowest possible accuracy!? Further, the lack of any physical discrimination between the information and the noise highlights the paradox by posing the question whether the Sun will rise tomorrow is also subject to prediction, the accuracy of which depends on the current entropy of the Solar system and its environment!? It is obvious that the core of these paradoxes lies in the parity between the information and the entropy established on the grounds of the oversimplified relation between order and disorder expressed through the dependence of both the information and the entropy on the probability for realization of a given state. In order to avoid the problem, we assert that the information must be defined so that to account for the functional organization of the "order". Since the functional organization of a given state is mapped onto the properties of the discrete band in the power spectrum, we proposed the latter to be the target measure of information. Note that any discrete band reveals the long-term correlations established in the system, i.e. they correspond to certain deterministic processes or causal relations. On the other hand, imposed on every stochastic process, the boundeness viewed as the only general constraint for providing the stability, makes the shape of the power spectrum in each realization insensitive to the particularities of its statistics [2, chapter 1]. In turn, the obtained insensitivity provides constant accuracy of the separation between the information (a discrete band) and the noise (a continuous band of universal shape). The major advantage of maintaining constant accuracy lies in meeting the most general requirement for the reproducibility of the information: it is worth noting that while the deterministic processes are reproducible by definition, the noise realizations are always different. For that reason the issue about the accuracy of the separation of the information from the noise appears as an indistinguishable part of the problem about the reproducibility in general.

## 5. Novel Approach to Security of Encrypting

The long-standing problem about the secure exchange of information becomes more and more serious in view of rapidly growing communications which use noisy public channels such as Internet, telephones etc. The problem has two major aspects that are particularly acute because of their immense social impact: on the one hand, it is necessary to diminish as much as possible the distortions of the information provoked by the contingencies of the traffic noise and on the other hand to protect it from intentional attacks on its privacy. So far, these problems have been treated predominantly separately. Though a remarkable success has been achieved and a number of ingenious strategies have been developed, the solution of these problems still lacks general approach, insensitive to the particularities of coding and ciphering (or their combination). The weak point of all developed so far strategies is the vulnerability of the encrypting protocol to the noise since the latter can essentially distort the encapsulated information. Yet, we assert that by means of the same protocol the information can be made both robust to the noise and secure to intentional attacks on its privacy. The solution lies in the properties of the proposed by us new definition of information – discrete band in the power spectrum of a bounded time series. As we have already proved in [3], that the general condition for providing long-term stability, the boundedness, ensures the insensitivity of the information to the noise statistics, i.e. whatever the statistics of the noise is, the information encapsulated in the discrete band can be obtained with the same accuracy over and over again. This result proves the robustness of the information to the noise in the general frame of boundedness as a single constraint imposed on the structure of the bounded time series. The next step is to demonstrate that the same circumstance helps for considerable enhancing of the robustness to intentional attacks on the privacy of the information. Indeed, by appropriate embedding of the information, viewed as a deterministic process, into bounded noise with arbitrary statistics we can produce a bounded irregular time series that can be safely transmitted trough public noisy channel. The great advantage of this approach is that the sender can change the statistics of the noise unilaterally and the receiver(s) still can decrypt the information encapsulated in the discrete band. The unilateral change of embedding noise statistics is a new powerful tool for fast reaction to any contingency and to any attempt to break the code and/or to compromise the key. The strategic benefit of the obtained unilaterality is that its use considerably reduces the need of permanent complicated multilateral synchronization of encryption and decryption keys persistent even with the use of asymmetric ones. In addition, it is worth noting that the "security through embedding into noise" is an entirely novel approach whose major property is that it is generic for all types of cryptographic protocols regardless to the particularities of the code, cipher and key used.

### Conclusions

The hallmark of the present paper is the proof that the thermal equilibrium is a universal dynamical process aimed to sustain the long-term stability of any complex system. This result substantiates the entirely new role of the thermal equilibrium and its understanding compared to that in the conventional thermodynamics. Note that the latter assumes it as a necessary prerequisite for establishing the equilibrium as the state of maximum entropy. This setting, however, is incompatible with any functional diversification since the latter needs spatial inhomogeneity which, however, does not allow reaching the maximum of the entropy. Indeed, reaching of the entropy maximum presupposes an insensitive to the partitioning additivity of the entropy. However, this requirement is met only for perfectly homogeneous systems. Thus, the new understanding of the thermal equilibrium is an essential step towards solving of the long-standing paradox why the thermodynamics explains so well the operation of the heat machines and fails with more complicated structures. Indeed, the reduction of the thermal equilibrium to a dynamical process that sustains the stability of a system regardless to its functional morphology releases it from the association with the state of maximum entropy viewed as the only admissible long-term stable state. Furthermore, the established in the paper strong link between the reproducibility and the long-term stability puts forward the boundedness as the general principle that provides stable and reproducible behavior of any complex system and thus calls for corresponding modyfication of the thermodynamics. In our recent paper [3] on constituting its framework we have demonstrated that the equilibrium is not unique; on the contrary, it can be any stable state established by the interplay between the concrete dynamics and the concrete functional organization. It is worth noting that now we prove that either of these states is in thermal equilibrium and thus we can call them "equilibriums". Moreover, transitions between these "equilibriums" are available in response to application of appropriate external stimuli. This result opens the door for the use of each of them as a logical unit on its input. Then, by means of applying appropriate stimuli, the "logical" landscape of a system can be modified so that to achieve desired logical output. Yet, it should be stressed that though in general the equilibrium is not anymore associated with the single state of the maximum entropy, the realization of this scenario is still possible but as a particular case. Thus, the fundamental property of the modified thermodynamics is that the launch of the ability to create information and its shutting down strongly depends on the functional organization of the system. In return, this property makes the reformulated thermodynamics equally appropriate for the heat machines and the living organisms.

### References

1. M. K. Koleva, http://arXiv.org/cond-mat/0212357
2. M. K. Koleva, http://arXiv.org/physics/0512078
3. M. K. Koleva, http://arXiv.org/physics/0609096
4. C. E. Shannon and W. Weaver, *The Mathematical Theory of Information*, University of Ill. Press, Urbana (1949)